\documentclass[useAMS,usenatbib,aas_macros,fleqn]{mn2e}

\usepackage{epsfig}
\usepackage{amssymb}
\usepackage{natbib}
\usepackage{aas_macros}
\usepackage{times}

\usepackage{amsmath}

\DeclareGraphicsExtensions{.eps}

\newcommand{\figu}{Figure~}
\newcommand{\figus}{Figures~}
\newcommand{\eq}{Equation~}

\newcommand{\sect}{Section~}

\newcommand{\BH}{black hole}
\newcommand{\BHs}{black holes}

\newcommand{\sis}{$\sigma$}
\newcommand{\sise}{\sigma}

\newcommand{\mbh}{$M_{\rm bh}$}
\newcommand{\mbhe}{M_{\rm bh}}

\newcommand{\re}{$R_{\rm e}$}
\newcommand{\ree}{R_{\rm e}}

\newcommand{\reb}{$R_{\rm e,bulge}$}
\newcommand{\rebe}{R_{\rm e,bulge}}

\newcommand{\mdyn}{$M_{\rm dyn}$}
\newcommand{\mdyne}{M_{\rm dyn}}

\newcommand{\mdynb}{$M_{\rm dyn, bulge}$}
\newcommand{\mdynbe}{M_{\rm dyn, bulge}}

\newcommand{\mstar}{$M_{\rm star}$}
\newcommand{\mstare}{M_{\rm star}}
\newcommand{\mbulge}{$M_{\rm bulge}$}
\newcommand{\mbulgee}{M_{\rm bulge}}

\newcommand{\nb}{$n_{\rm sph}$}
\newcommand{\nbe}{n_{\rm sph}}

\newcommand{\msune}{M_{\odot}}

\newcommand{\SerExp}{\textsc{SerExp}}

\newcommand{\modI}{Model I}

\newcommand{\papI}{Paper I}

\begin{document}

\def\sarc{$^{\prime\prime}\!\!.$}
\def\arcsec{$^{\prime\prime}$}
\def\arcmin{$^{\prime}$}
\def\degr{$^{\circ}$}
\def\seco{$^{\rm s}\!\!.$}
\def\ls{\lower 2pt \hbox{$\;\scriptscriptstyle \buildrel<\over\sim\;$}}
\def\gs{\lower 2pt \hbox{$\;\scriptscriptstyle \buildrel>\over\sim\;$}}

\title[SMBHs: selection bias and S\'{e}rsic indices]{Selection bias in dynamically-measured super-massive black hole samples: dynamical masses and dependence on S\'{e}rsic index}

\author[F. Shankar et al.]
{Francesco Shankar$^{1}$\thanks{E-mail:$\;$F.Shankar@soton.ac.uk},
Mariangela Bernardi$^{2}$, Ravi K. Sheth$^{2}$
\\
$1$ Department of Physics and Astronomy, University of Southampton, Highfield, SO17 1BJ, UK\\
$2$ Department of Physics and Astronomy, University of Pennsylvania, 209 South 33rd St, Philadelphia, PA 19104\\
}

\date{}
\pagerange{\pageref{firstpage}--
\pageref{lastpage}} \pubyear{2016}
\maketitle
\label{firstpage}

\begin{abstract}
We extend the comparison between the set of local galaxies having dynamically measured \BHs\ with galaxies in the Sloan Digital Sky Survey (SDSS). We first show that the most up-to-date local \BH\ samples of early-type galaxies with measurements of effective radii, luminosities, and S\'{e}rsic indices of the bulges of their host galaxies, have dynamical mass and S\'{e}rsic index distributions consistent with those of SDSS early-type galaxies of similar bulge stellar mass. The host galaxies of local \BH\ samples thus do not appear structurally different from SDSS galaxies, sharing similar dynamical masses, light profiles and light distributions. Analysis of the residuals reveals that velocity dispersion is more fundamental than S\'{e}rsic index \nb\ in the scaling relations between black holes and galaxies.
Indeed, residuals with \nb\ could be ascribed to the (weak) correlation with bulge mass or even velocity dispersion.
Finally, targetted Monte Carlo simulations that include the effects of the sphere of influence of the black hole, and tuned to reproduce the observed residuals and scaling relations in terms of velocity dispersion and stellar mass, show that, at least for galaxies with $\mbulgee \gtrsim 10^{10}\, \msune$ and $\nbe\gtrsim 5$, the observed mean black hole mass at fixed S\'{e}rsic index is biased significantly higher than the intrinsic value.
\end{abstract}


\section{Introduction}
\label{sec|intro}

The scaling relations between supermassive black holes and their host galaxies have been a very hot topic in the last thirty years \citep[see, e.g.,][for reviews]{FerrareseFord,ShankarReview,GrahamReview15}. This is because such scalings may be the smoking gun of a ``co-evolution'' between the two systems \citep[e.g.,][]{SilkReviewGalx}, although the physical processes involved are still highly debated, ranging from quasar feedback to black hole mergers, clumpy accretion, and/or galaxy-scale gravitational torques \citep[e.g.,][]{SilkRees,Vittorini05,JahnkeMaccio,Bournaud11b,Alcazar15}. Besides the well-known correlations with velocity dispersion \sis\ \citep[][]{Ferrarese00,Gebhardt00} and (bulge) stellar mass \mbulge\ \citep[e.g.,][]{Marconi03,Lauer07demo,KormendyHo,Laesker14,Saglia16},
correlations with the light concentration and S\'{e}rsic index have also been measured \citep[e.g.,][and references therein]{Graham01,GrahamDriver,Savo16n}.

The correlation between black hole mass and S\'{e}rsic index, in particular, has been the subject of numerous studies in recent years. Some groups \citep[e.g.,][]{Sani11,Beifiori12} have not detected any significant correlation, while more recently \citet{Savo16n}, by compiling a larger galaxy sample with accurate and uniform photometric decompositions, has claimed a significant correlation characterized by a slope of $3.39\pm0.15$ and an intrinsic scatter of $\sim 0.25$ dex.  The scatter is comparable to, or even smaller than, the one in the scaling with velocity dispersion, paving the way for its use as a black hole mass indicator in galaxies \citep[e.g.,][]{Graham07BHMF,Pakdil16}.

Unveiling the actual existence of the black hole-S\'{e}rsic index relation could be a key piece of evidence for some important galaxy evolutionary patterns. For example, more or less violent disc instabilities in gas-rich, high-redshift discs could feed both an inner bulge and a central black hole \citep[e.g.,][]{Bournaud11a}. A progressively more prominent bulge component, possibly characterized by a proportionally increasing galaxy S\'{e}rsic index, may then be able to halt star formation in the host galaxy \citep[e.g.,][]{Martig09,DekelBurkert}. An initial correlation between black hole mass and S\'{e}rsic index could have thus been established by these high-redshift dissipative processes. If galaxy mergers have been the actual drivers behind the origin of the large sizes and high S\'{e}rsic indices in present-day massive galaxies \citep[e.g.,][]{Hilz13,Nipoti15}, then black holes should have necessarily followed in some degree their host galaxy mergers to preserve a correlation with S\'{e}rsic index.

On the other hand, both disc instabilities and repeated black hole mergers should also induce the build-up of a closer link between black hole mass and stellar mass \citep[e.g.,][]{JahnkeMaccio}, at variance with the recent results by our group \citep[][\papI\ hereafter]{Shankar16BH} and others \citep[][]{Bluck16,Remco16}. In \papI we showed that, following a number of previous claims \citep[e.g.,][]{Bernardi07,Remco15}, the local sample of galaxies with dynamically-measured supermassive black holes is highly biased with respect to an unbiased large sample of galaxies of similar stellar mass. In particular, black hole galactic hosts appear to have significantly higher velocity dispersion (and slightly lower sizes) at fixed stellar mass. \papI\ used Monte Carlo simulations and residual analysis to show that such biases can result if the sample of local galaxies is preselected with the requirement that the black hole sphere of influence must be resolved to measure black hole masses with spatially resolved kinematics. The same simulations and statistical analysis clearly point to velocity dispersion being more fundamental than stellar mass or effective radius, and predict significantly lower normalizations for the intrinsic scaling relations.  The latter partly solves the systematic discrepancy between dynamically-based black hole-galaxy scaling relations versus those of active galaxies \citep[e.g.,][]{ReinesVolonteri15}, favouring proportionally lower virial calibration factors $f_{vir}$ for estimating black hole masses in active galaxies \citep[e.g.,][]{HoKim14}.

However, it is possible that some of the bias
may be induced by real \emph{structural} differences,
i.e., physical effects could also be playing a role. One of the two aims of this Letter is to address the question of structural differences between local galaxies with dynamically-measured black holes and their counterparts in large unbiased samples of galaxies. After briefly introducing the data adopted in this work in \sect\ref{sec|data}, we focus on dynamical masses and (bulge) S\'{e}rsic \nb\ distributions in \sect\ref{sec|Results}.  We then move to the second aim of this work, which is to compare the importance of S\'{e}rsic index with other variables in the black hole scaling relations, in order to determine if \nb\ plays a fundamental role.  We use dedicated Monte Carlo simulations to interpret our results and present our conclusions in \sect\ref{sec|discu}.  Two Appendices provide details of our analysis.  Appendix~A describes how our analysis accounts for statistical measurement errors, and Appendix~B shows how the slopes of correlations involving three variables are related to slopes of pairwise regressions.

\section{Data}
\label{sec|data}

\begin{figure*}
    \center{\includegraphics[width=17truecm]{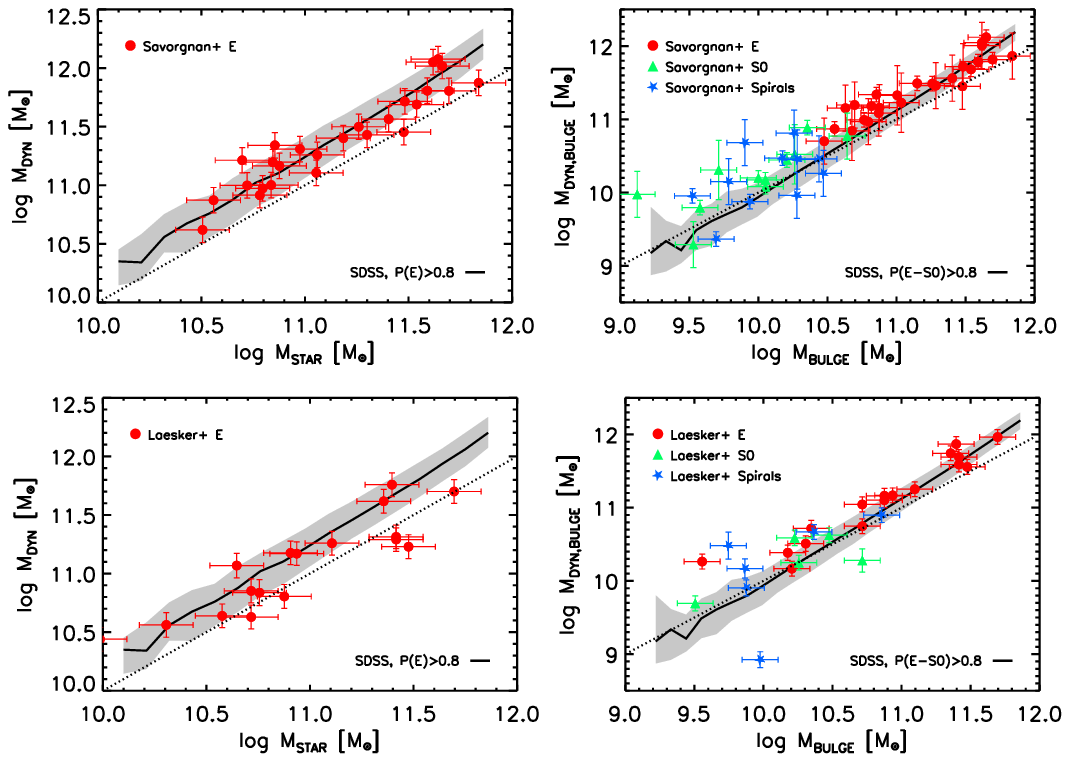}
    \caption{Left: Mean dynamical mass, $\mdyne=K(\nbe)\ree\sise^2/G$, as a function of stellar mass.  Right: Same format as the left panels but for the bulge component:  $\mdynbe=K(\nbe) \rebe\sise^2/G$ as a function of \mbulge. Solid lines in each panel show the mean relation defined by the SDSS of only E (left) or E-S0 (right) samples, with the \SerExp\ stellar masses and photometric parameters from \citet{Meert15}; grey bands mark the $1\sigma$ dispersion around the mean.
    Symbols show the \citet[][top panels]{Savo15} and \citet[][bottom panels]{Laesker14} samples. Filled red circles, green triangles, and blue stars show, respectively, ellipticals, lenticulars, and spirals, the latter two reported only in the right panels. Dotted lines in each panel mark the one-to-one relations. The agreement with the SDSS galaxies is good.
    \label{fig|Mdyn}}}
\end{figure*}

Following \papI, we use the \citet{Savo15} sample of galaxies having dynamically measured \BHs, with self-consistent\footnote{The same surface brightness profile fitting procedure has been adopted for each of the 66 galaxies in the sample.} measurements of S\'{e}rsic luminosities, effective radii and S\'{e}rsic indices of the spheroidal components, as well as estimates of the total host galaxy luminosities and effective radii. Central velocity dispersions are from Hyperleda, while stellar masses are obtained by applying to the 3.6$\mu$m (Spitzer) luminosities a constant mass-to-light ratio of $(M/M_\odot)/(L/L_\odot)=0.6$ from \citet[e.g.][]{Meidt14}.  As detailed in \papI, from the original sample of 66 galaxies we remove 18 objects with uncertain black hole mass and/or surface brightness, or unavailable central velocity dispersion, or because they are ongoing mergers. We checked that our results are not affected by the removal of these sources. The errors quoted by \citet{Savo15} on the photometric parameters include systematics (e.g., from comparison with different authors and analysis methods).  However, since we will be interested in scaling relations - the estimate of which includes accounting for errors -  we do not include the systematic contribution to the error on \nb\ at this point.  Specifically, we only account for random errors when estimating the intrinsic slope, zero-point and scatter.  We assess the influence of systematics as follows.  When a different analysis method is used to estimate the photometric parameters, then we use these new values to estimate scaling relations in the same way as before (i.e., accounting only for the random errors associated with these new values).  The differences between the inferred scaling relations contribute to the systematic error on the inferred scaling relation. In practice, we used as the ``other values'' the sample of \citet{Laesker14}, which also includes accurate photometric analysis from the WIRcam imager at the CanadaFranceHawaiiTelescope, with S\'{e}rsic-based light profile fitting routines. We retain 28 galaxies from their original sample, containing the most secure dynamical black hole mass measurements according to \citet{KormendyHo}.

To represent the full galaxy sample, we use objects in the Sloan Digital Sky Survey (SDSS) DR7 spectroscopic sample \citep{2009ApJS..182..543A} in the redshift range $0.05<z<0.2$ with the photometric measurements from \citealt[][]{Meert15}.  Throughout this paper, we restrict the analysis to galaxies whose probability of being elliptical or lenticular $p$(E--S0) is greater than $0.80$, based on the Bayesian automated morphological classifier by \citet{Huertas11}; we refer to this as the SDSS E-S0 sample\footnote{Following \papI, when dealing with bulges we preferentially adopt E-S0 galaxies with $p$(E--S0)$>0.8$ as our reference SDSS comparison sample, because determining the central velocity dispersion of spirals from the SDSS spectra (which are not spatially resolved) is not possible. We checked, however, that none of our results depends on the exact cut in $p$(E--S0).}. When dealing with total stellar masses we will instead refer to only ellipticals with $p$(E)$>0.8$. Stellar masses are derived by combining the \SerExp\ estimates of the luminosity from \citet{Meert15} with
mass-to-light ratios $\mstare/L$
detailed in \citet{Bernardi10,Bernardi13} and \citet{Chabrier03} Initial Mass Function (IMF).  Systematic differences in $\mstare/L$ can be of order 0.1~dex \citep[e.g.][]{Bernardi16ML}.
SDSS velocity dispersions are converted from $R_e/8$ to the 0.595 kpc aperture of the Hyperleda \footnote{From here onwards, unless otherwise stated, velocity dispersions \sis\ will always be defined at the aperture of Hyperleda.} database \citep{Paturel03}, using the mean aperture corrections \citep[e.g.,][]{Jorgensen96,Cappellari06}
\begin{equation}
\left(\frac{\sigma_R}{\sigma_e} \right) = \left(R/R_e\right)^{-0.066}\, .
\label{eq|Cappellari}
\end{equation}
We note that blurring by seeing effects could potentially reduce central velocity dispersion measurements \citep[e.g.,][]{Graham98}, however we do not foresee any major difference in the seeing affecting ground-based measurements of \sis\ in SDSS and those catalogued in Hyperleda. Strictly speaking, the S\'{e}rsic index \nb\ we will adopt in this work is always referred to the galaxy \emph{spheroidal} component extracted from a \SerExp\ luminosity profile fitting in both the SDSS and the \citet{Savo15} and \citet{Laesker14} samples. The half-light radii \reb\ and \re\ are defined as the radii containing half of the bulge and total galaxy luminosity, respectively. In the following, we will label the total galaxy stellar mass, galaxy bulge stellar mass, total galaxy dynamical mass, and galaxy bulge dynamical mass as \mstar, \mbulge, \mdyn, and \mdynb, respectively. In the next sections, unless otherwise noted, we will compute median instead of mean quantities. While this makes little difference when dealing with stellar/dynamical masses or velocity dispersions, it matters more with the (non-Gaussian) S\'{e}rsic distributions at fixed stellar mass, for which medians are more appropriate.

\section{Results}
\label{sec|Results}

\begin{figure*}
    \center{\includegraphics[width=17truecm]{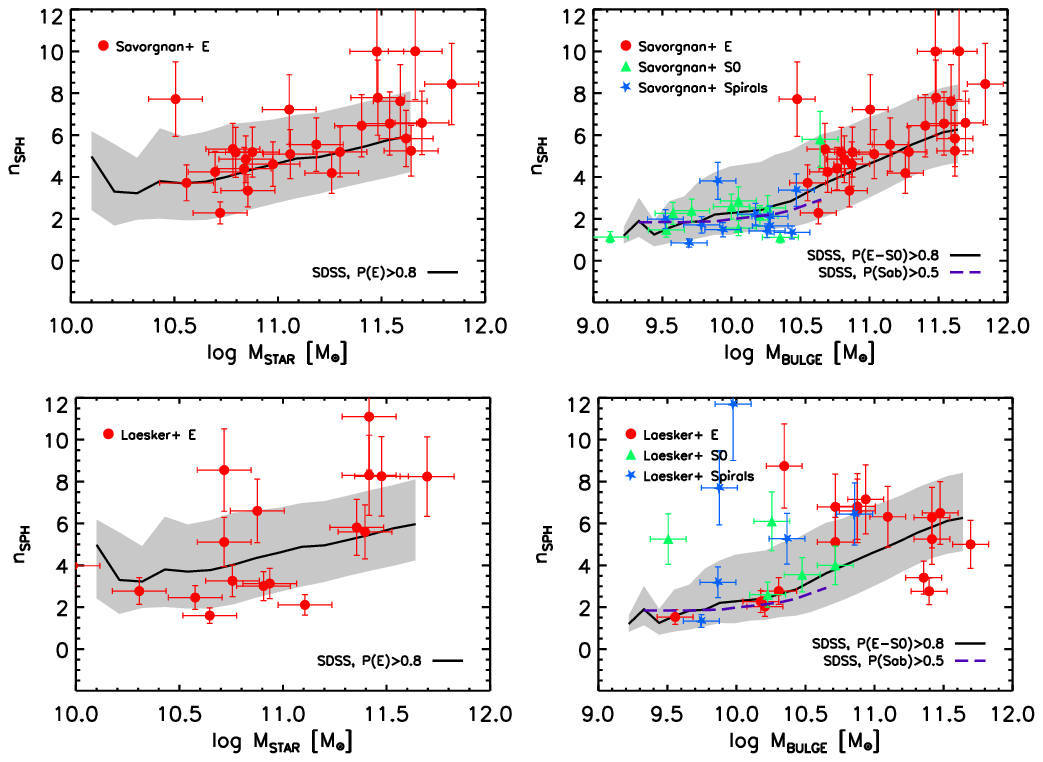}
    \caption{S\'{e}rsic index \nb\ as a function of galaxy total stellar mass (left) and bulge stellar mass (right).  Symbols show the \citet[][top panels]{Savo15} and \citet[][bottom panels]{Laesker14} samples, divided into ellipticals, lenticulars, and spirals, as labelled.  Solid line and grey shaded region show the relation defined by SDSS only E (left) or E-S0 (right) samples (black lines with grey areas). The purple long-dashed line in the right panels shows the median S\'{e}rsic index for SDSS Sab galaxies. There is no significant mismatch between SDSS galaxies and black hole samples.
    \label{fig|Sersic}}}
\end{figure*}

To test the hypothesis that galaxies with dynamically-measured black holes are a structurally different subset of the full galaxy population -- represented by the SDSS -- \figu\ref{fig|Mdyn} shows the mean dynamical mass (solid lines), along with its 1$\sigma$ dispersion (grey bands), for the SDSS E-S0 galaxies as a function of total (left panels) and bulge (right panels) stellar mass. The SDSS E-S0s are compared to the \citet{Savo15} and \citet{Laesker14} samples (top and bottom panels, respectively), divided into ellipticals, lenticulars/S0, and spirals, as labelled. Here dynamical mass is always computed for both samples as $\mdyne=K(\nbe)\ree\sise^2/G$, with the S\'{e}rsic index-dependent virial constant $K(\nbe)$ taken from \citet{Prugniel97}. It is clear that the bulge dynamical mass of all galaxy types in the \citet{Savo15} and \citet{Laesker14} samples broadly agree with those of SDSS E-S0s galaxies of similar stellar mass. The data tend to show slightly larger dynamical bulge masses at lower stellar bulge masses (right panels in \figu\ref{fig|Mdyn}), most probably induced by the very large velocity dispersions characterizing the low mass galaxies with dynamical measurements of black holes, as emphasized in \papI. However, most of the \citet{Savo15} and \citet{Laesker14} data are still broadly consistent with SDSS galaxies within the quoted uncertainties. In line with a number of previous studies \citep[e.g.,][and references therein]{Forbes08,ShankarBernardi09,Bernardi11b,Cappellari13}, it is also interesting to note that in both the SDSS and \citet{Savo15} samples all ellipticals have a dynamical mass a factor of $\sim 2$ higher than their total stellar mass (left); this ratio is smaller but still greater than unity if only the bulge component is used (right; compare solid and dotted lines, the latter marking the one-to-one relations).

\figu\ref{fig|Sersic} shows the correlation between S\'{e}rsic \nb\ and total (left) or bulge (right) stellar mass. Solid lines and grey regions mark the median and 1$\sigma$ dispersions for the SDSS only E (left) or E-S0s (right).  Symbols show the \mbh\ hosts from \citet[][top panels]{Savo15} and \citet[][bottom panels]{Laesker14}.  The panels on the left show that ellipticals (red circles) match the SDSS S\'{e}rsic index distributions. The match is extended to lenticulars (green triangles) when switching to bulge stellar masses (right panel). Spirals (blue stars) in the \citet{Savo15} sample (top, right) tend to fall slightly below the median traced by the SDSS E-S0 galaxies, but are within the median S\'{e}rsic distributions of E-S0 and consistent with the Sab (purple long-dashed line) SDSS galaxies.  Thus, the top panels of \figu\ref{fig|Sersic} suggest that local galaxies with black hole mass measurements are \emph{not}, on average, structurally different from SDSS galaxies of similar stellar mass.

The bottom panels show a similar analysis of the \citet{Laesker14} sample.  In both panels, the correlations are much noisier than before.  Spirals tend to lie somewhat above the median SDSS S\'{e}rsic index of SDSS galaxies. In fact, the symbols in the bottom right panel suggest that \nb\ decreases as \mbulge\ increases; this is opposite to the trend in the \citet{Savo15} sample, and will be important in what follows.  This difference shows how challenging accurate determinations of S\'{e}rsic indices can be. Finally, we also verified that, for the early-type galaxies in our black hole mass samples, the projected mass density within a few kpc are similar to, if not lower than those of SDSS galaxies of similar bulge mass or velocity dispersion.

\begin{figure*}
    \center{\includegraphics[width=17truecm]{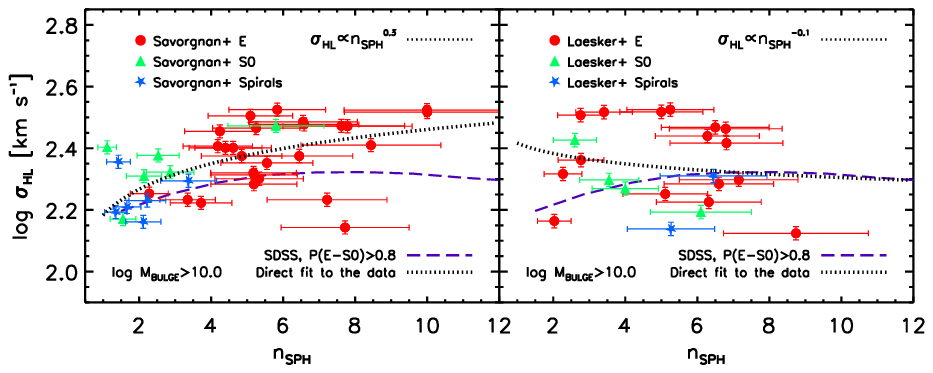}
    \caption{Correlation between velocity dispersion and S\'{e}rsic index \nb. Long-dashed purple line is the median relation in SDSS early-type galaxies, while symbols mark the galaxies in the \citet[][left panel]{Savo15} and \citet[][right panel]{Laesker14} samples having $\log \mstare/\msune >10$, divided per morphological type, as labelled. The black thick dotted lines are the direct fits to these data. The \citet[][left panel]{Savo15} sample, in particular, has a higher normalization and a steeper slope than the SDSS relation.
    \label{fig|SigmaSersic}}}
\end{figure*}

\figu\ref{fig|SigmaSersic} shows that the mean velocity dispersion as a function of S\'{e}rsic index \nb\ for early-type galaxies in our SDSS sample (long-dashed purple line) is rather flat\footnote{In contrast, the mean S\'{e}rsic index is a steeper function of velocity dispersion, though the scatter is large.} at $\nbe \gtrsim 5$. A direct fit to the data by \citet{Savo15}, reported in the left panel of \figu\ref{fig|SigmaSersic} and labelled per morphological type, yields a systematically higher and steeper correlation with $\sise\propto \nbe^{0.3}$ (black thick dotted line).  We interpret this as another sign of existing biases in the local sample of galaxies with dynamical measurements of black holes, in line with \papI. The \citet{Laesker14} sample instead (right panel of \figu\ref{fig|SigmaSersic}) appears broadly consistent with SDSS data, with a negligible dependence on S\'{e}rsic index, especially at high \nb, as in our SDSS data. In \figu\ref{fig|SigmaSersic} we only show galaxies with $\log \mbulgee/\msune >10$, to make a fair comparison with our (selection biased) SDSS E-S0 mock sample, described in the next section, which can reliably probe only above this lower limit in bulge mass.

\begin{figure*}
    \center{\includegraphics[width=17truecm]{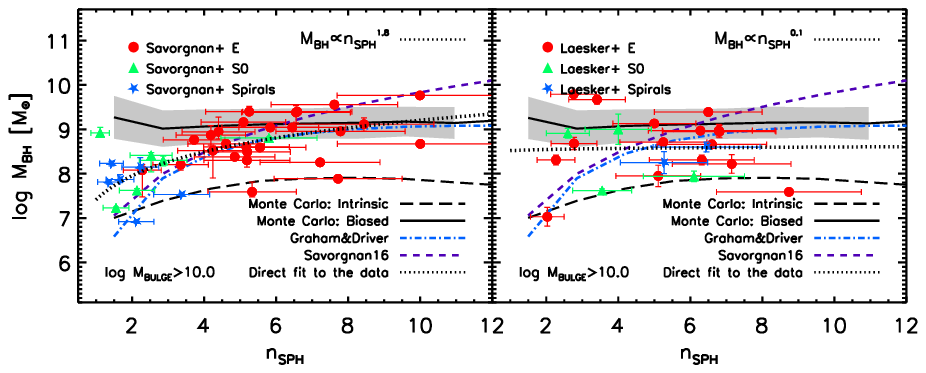}
    \caption{Correlation between black hole mass \mbh\ and bulge S\'{e}rsic index \nb.  Symbols show the galaxies in the \citet[][left panel]{Savo15} and \citet[][right panel]{Laesker14} samples having $\log \mstare/\msune >10$.  Blue dot-dashed and purple dotted lines are the curved relations described by \citet{GrahamDriver} and \citet{Savo16n}, respectively.  Black solid lines and grey bands show the selection biased relation in the Monte Carlo simulations described in the next section when the intrinsic relation is given by \modI\ of \citet{Shankar16BH} (dashed black line).  This (selection biased relation) is broadly similar to that observed, suggesting that the mean black hole mass at fixed \nb\ can be severely overestimated, at least for $\nbe\gtrsim 5$. The black thick dotted lines are the direct fits to the data. The \citet[][right panel]{Laesker14} sample, in particular, shows no dependence on S\'{e}rsic index and it is broadly in line with the predictions of the Monte Carlo simulations.}
    \label{fig|Model}}
\end{figure*}

\figu\ref{fig|Model} shows the correlation between black hole mass \mbh\ and bulge S\'{e}rsic index \nb.  Symbols show the galaxies in the \citet[][]{Savo15} and \citet[][]{Laesker14} samples (left and right panels respectively) having $\log \mstare/\msune >10$.  Blue dot-dashed and purple dotted lines are the curved relations described by \citet{GrahamDriver} and \citet{Savo16n}, respectively. We describe the grey regions and other curves later. A direct fit to the \citet{Savo15} and \citet{Laesker14} data yields $\mbhe\propto\nbe^{1.8}$ and $\mbhe\propto\nbe^{0.1}$, respectively. The Appendix addresses the question of whether or not such (different) behaviours would be expected  if black hole mass is closely correlated with velocity dispersion, as emphasized in \papI, but the \sis-\nb\ trends for the two samples are ery different (as shown in \figu\ref{fig|SigmaSersic}).

For this purpose, we now test if the correlation between black hole mass and S\'{e}rsic index, evident at least in the \citet{Savo15} sample, is fundamental, or merely a consequence of others.  Correlations between the residuals of scaling relations are an efficient way of addressing this question \cite[][and \papI]{ShethBernardi12}.

The original errors assigned to the \citet{Savo16n} sample include both the statistical as well as the systematic errors that affect photometric decompositions. This is a particularly relevant issue for S\'{e}rsic indices. The quoted errors in \nb\ are in fact of the order of $\sim 35\%$, while typical statistical errors amount to at most $\lesssim 20-25\%$, i.e., $\lesssim$ 0.1 dex \citep{Bernardi12}. As discussed in \sect\ref{sec|data}, when computing residuals with respect to \nb, we will always consider only the \emph{statistical} $\sim 0.1$~dex errors. The difference in the measured slopes from different samples should then provide an indication of the impact of additional systematic uncertainties.  We note that the impact of systematic uncertainties should not be included in any single measurement simply by inflating the measured statistical uncertainties. For similar reasons we adopt typical average errors for the bulge stellar masses of 0.13 dex, i.e., 30\% \citep[see, e.g.,][]{Meert13}, instead of their reported average value of $\sim 0.17$ dex.  Appendix~A describes in some detail how we account for statistical measurement errors, and assign error bars in the analysis which follows.

\begin{figure*}
    \center{\includegraphics[width=17truecm]{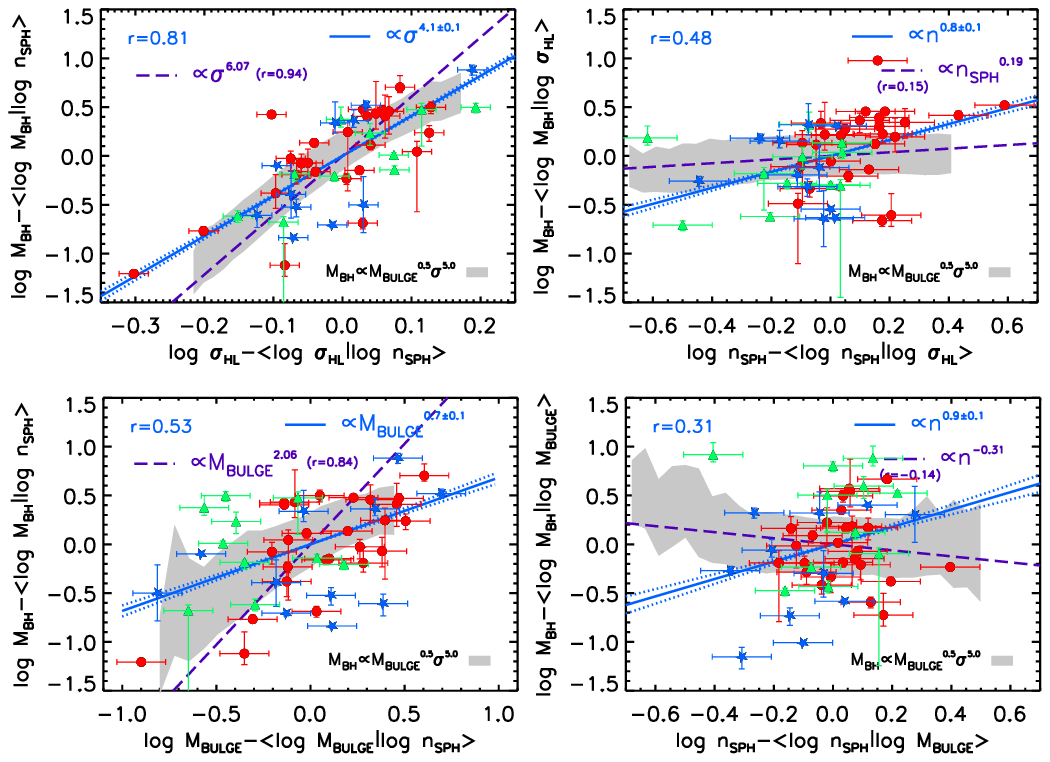}
    \caption{Correlations between residuals from the observed scaling relations, as indicated in each panel.  Red circles, green triangles, and blue stars show ellipticals, lenticulars, and spiral galaxies in the \citet{Savo15} sample. The blue solid and dotted lines mark the best-fit scaling relation and the 1$\sigma$ uncertainty in the slope (best-fit slopes are reported in the upper, right corners). The Pearson correlation coefficient $r$ is reported in the top, left corner of each panel. The grey bands and purple long-dashed lines show the residuals extracted from the Monte Carlo simulations described in the text with and without selection in the black hole gravitational sphere of influence. The residual correlations with S\'{e}rsic index at fixed velocity dispersion (top right panel) and, especially, with (bulge) stellar mass (bottom right panel), are weak.
    \label{fig|Residuals}}}
\end{figure*}

The upper left panel of \figu\ref{fig|Residuals} shows that residuals in the \citet{Savo16n} sample from the \mbh-\nb\ relation correlate very well with those from the \sis-\nb\ relation:  the Pearson coefficient is $r=0.81$. In contrast, the upper right panel shows that residuals from the \mbh-\sis\ relation show a much weaker correlation with those from the \nb-\sis\ correlation ($r=0.48$).  Together, the two upper panels imply $\mbhe \propto \sise^{4.1\pm0.1} \,\nbe^{0.8\pm0.1}$.

Similarly, the two lower panels imply $\mbhe \propto \mstare^{0.7\pm0.1} \,\nbe^{0.9\pm0.1}$.  However, the correlation with bulge mass at fixed \nb\ (lower left panel) tends to be tighter than the one in S\'{e}rsic index at fixed \mbulge\ (lower right panel has $r\lesssim 0.31$).  Both slope and Pearson correlation coefficient drop to about zero when considering only E-S0 galaxies, suggesting that most of the correlation in \figu\ref{fig|Model} between black hole mass and S\'{e}rsic index could be induced by the relation between S\'{e}rsic index and stellar (bulge) mass. If barred galaxies are excluded from the \citet{Savo15} sample, then the Pearson coefficients in the two right hand panels of \figu\ref{fig|Residuals} decrease to $r\sim 0.33$ (top) and $r\sim 0.14$ (bottom). Our analysis thus strongly suggests that velocity dispersion is more fundamental than S\'{e}rsic index, further supporting and extending the results in \papI.

\begin{figure*}
    \center{\includegraphics[width=17truecm]{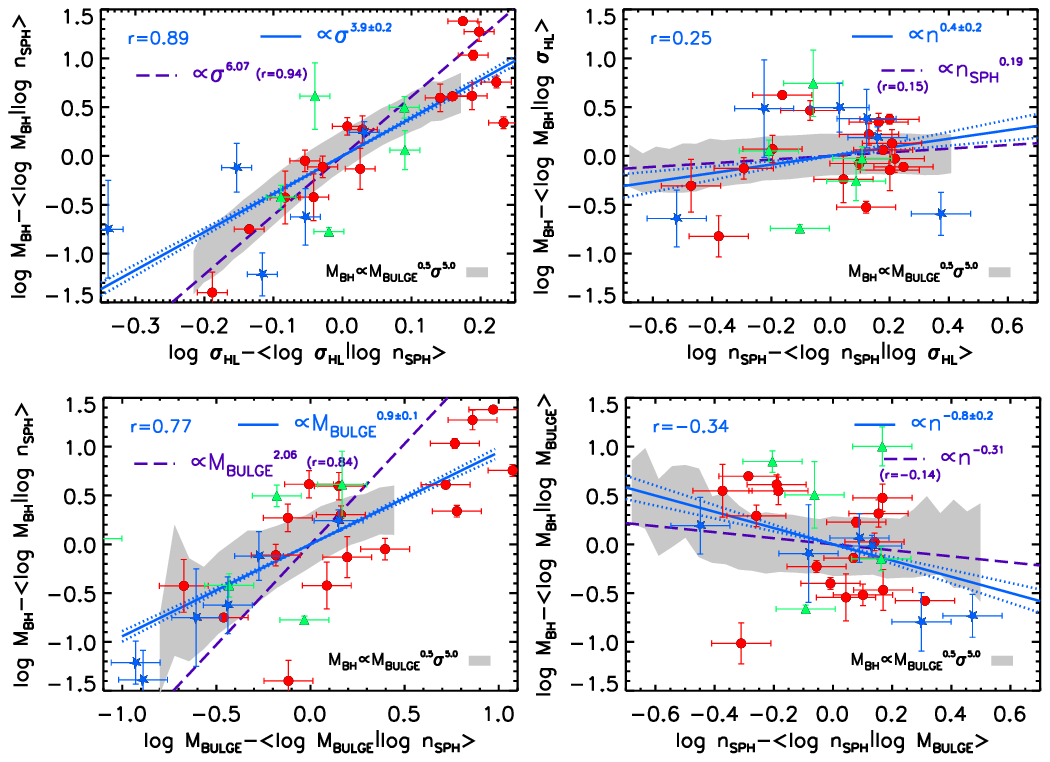}
    \caption{Same as \figu\ref{fig|Residuals} but for the \citet{Laesker14} sample. The residual correlations with S\'{e}rsic index at fixed velocity dispersion and stellar mass are extremely weak.
    \label{fig|ResidualsLaesker}}}
\end{figure*}

A similar analysis of the \citet{Laesker14} sample, reported in \figu\ref{fig|ResidualsLaesker}, also yields a tight correlation with velocity dispersion ($r=0.89$ in upper left panel), and extremely weak correlations with S\'{e}rsic index ($r<0.3$ in top and bottom right panels). Using only E-S0 galaxies yields even stronger dependence on velocity dispersion and nearly no dependence on S\'{e}rsic index. Even assuming substantially larger statistical uncertainties in \nb\ still yields very weak correlations in the panels on the right.  Finally, note that \citet{Laesker14} also provide S\'{e}rsic indices derived allowing for a core in some galaxies \citep[see][for details]{Laesker14}.  Using these instead yields results consistent with \figu\ref{fig|ResidualsLaesker}.

In the analyses above, the errors on velocity dispersions were taken to be 5\% \citep[e.g.,][]{Tremaine02,Graham13}, in line with what is quoted in the Hyperleda data base. However, larger errors in velocity dispersion for these same galaxies have been reported in the literature \citep[e.g.,][]{Ferrarese02demo}, in line with those measured for SDSS galaxies \citep[e.g.,][]{Bernardi11a}. Larger errors in velocity dispersion would strengthen our main result that velocity dispersion is more fundamental than S\'{e}rsic index.

\begin{table}
	\centering
	\caption{Slopes of linear relations in our SDSS galaxy sample.}
	\label{tab:SDSSslopes}
	\begin{tabular}{ccccc} 
		\hline
        &  & &  X & \\
        \hline
		& & $\log \mstare$ & $\log \sise$ & $\log \nbe$ \\
		\hline
	  &	$\log \mstare$ &      & 2.05 & 0.36 \\
	Y &	$\log \sise$   & 0.33 &      & 0.16 \\
	  &	$\log \nbe$  & 0.19 & 0.55 &      \\
		\hline
	\end{tabular}
\end{table}

\section{Discussion}
\label{sec|discu}

In the previous section we showed that velocity dispersion is more fundamental than S\'{e}rsic index \nb\ for determining \mbh. Indeed, the \mbh-\nb\ correlation seems to be mostly induced by the \nb-\mbulge\ and \mbh-\mbulge\ relations. However, because the \mbh\ sample is biased (to large \sis) by the way in which the sample was selected, we must make sure that the relations defined by the symbols in \figu\ref{fig|Residuals} are not affected by the selection effect.  We use targetted Monte-Carlo simulations to do so:  details are given in \papI, so here we briefly summarize the main points.

To each SDSS galaxy in our sample\footnote{The simulations are based on the SDSS sample from \citet{Meert13} which is magnitude limited, though all mock residuals are weighted through $V_{\rm max}$. We have further verified that none of our conclusions are changed if we adopted a full mock case extracted from the stellar mass function and to which velocity dispersions, bulge fractions and S\'{e}rsic indices are assigned via empirically-based correlations.} we associate a supermassive black hole following the favoured model in \papI\
\begin{equation}
   \log \frac{\mbhe}{\msune}=
   \gamma + \beta \log\left(\frac{\sise}{200\, {\rm km\, s^{-1}}}\right) + \alpha\log\left(\frac{\mbulgee}{10^{11}\, \msune}\right)\, ,
   \label{eq|MbhSigma}
\end{equation}
with $(\gamma,\beta,\alpha)=(7.7,5.0,0.5)$ and a total (Gaussian) scatter of 0.25 dex (inclusive of observational errors).
We repeat the above procedure several times to create a ``full'' \BH\ sample, and retain only those objects for which the gravitational sphere of influence is greater than the typical resolution of the Hubble Space Telescope, i.e., $r_{\rm infl}\equiv G\mbhe/\sise^2>0.1''$.

First, we note that the selection-biased mock residuals predicted by our Monte Carlos (gray bands in \figus\ref{fig|Residuals} and \ref{fig|ResidualsLaesker}), predict strong correlations, especially in velocity dispersion, at fixed S\'{e}rsic index (left panels), and weak correlations with S\'{e}rsic index, in agreement with the \citet{Laesker14} sample, but not with the \citet{Savo15} one. It is interesting to note that the predictions of the Monte Carlos \emph{without} selection bias (purple long dashed lines) would predict significantly steeper residuals at fixed S\'{e}rsic index (see the Appendix for further details).

The long-dashed black lines in \figu\ref{fig|Model} shows the intrinsic \mbh-\nb\ relation in our SDSS E-S0 sample predicted by \eq\ref{eq|MbhSigma}.  It is remarkably flat, because velocity dispersion is a weak function of S\'{e}rsic index (\figu\ref{fig|SigmaSersic}; see Appendix for more discussion.) The solid black line and associated grey region show the mean and $1\sigma$ dispersion in the predicted \mbh-\nb\ relation of the selection biased sample (i.e., after selecting objects with large enough $r_{\rm infl}$).  Notice that it lies almost an order of magnitude above the intrinsic relation at $\nbe\gtrsim 5$.

For completeness, blue dot-dashed and purple dotted lines in \figu\ref{fig|Model} show fits to the observed \mbh-\nb\ relation from \citet{GrahamDriver} and \citet{Savo16n}, respectively. At least for relatively massive, large \nb\ early-type galaxies, these fits and the measurements are in broad agreement with the grey region defined by our selection-biased Monte Carlos. Hence, we conclude that at least some of the difference between the intrinsic relation (black long-dashed line) and the data at large \nb\ can be ascribed to selection effects.

At smaller \nb\ and lower \mstar\ the data by \citet{Savo15} tend to curve downwards as indicated by the fits, whereas our Monte Carlos do not. Including an intrinsic dependence between \mbh\ and \nb, despite not being favoured by the residuals in \figu\ref{fig|Residuals}, still produces a flat biased \mbh-\nb\ relation.  It may be that other, possibly mass-dependent, selection effects should be included in our Monte Carlos to account for the S\'{e}rsic index distribution of the $\log \mbulgee/\msune \lesssim 10$ galaxies in the local samples of galaxies with dynamically measured black holes. See the Appendix for further discussion of the expected slopes of the grey regions in \figus4--6.

To summarize, in this work we have compared SDSS early-type galaxies with the local sample of galaxies with dynamically-measured black holes from the \citet{Savo15} and \citet{Laesker14} samples with self-consistent estimates of bulge luminosities, effective radii, and S\'{e}rsic indices. We find the latter sample to be consistent with SDSS galaxies in terms of dynamical mass and S\'{e}rsic index distributions. Analysis of the residuals in \figus\ref{fig|Residuals} and \ref{fig|ResidualsLaesker}, reveals that velocity dispersion is more fundamental than S\'{e}rsic index \nb\ in the scaling relations between black holes and galaxies. Indeed, residuals with \nb\ could be ascribed to the underlying correlations with \sis\ and \mbulge.  Our conclusions are supported by targetted Monte Carlo tests that include the effects of the sphere of influence of the black hole. They show that, at least for galaxies with $\mbulgee \gtrsim 10^{10}\, \msune$ and $\nbe\gtrsim 5$, the observed median black hole at a given \nb\ is biased higher than the intrinsic value by up to an order of magnitude, i.e., black hole masses are over-predicted at the high-mass end, as was also revealed for the \mbh-\mbulge\ and \mbh-\sis\ relations (\papI).

\section*{Acknowledgments}
We warmly thank Alister Graham, Giulia Savorgnan, and Ronald L\"{a}sker for providing their data in electronic format and for useful discussions.

\appendix
\section{Accounting for measurement errors}
To include errors in the determination of the correlations, especially those between residuals, we follow \citet{Bernardi03} and \citet{ShethBernardi12}. For any set of measurements ${x_i,y_i}$ and (normalized) weights $w_i$, we first compute the linear relations with slope $m_{\rm y|x}$ and zero point $zp_{\rm y|x}$ given by
\begin{equation}
m_{\rm y|x}=\frac{S_{xy}-E_{xy}}{S_{xx}-E_{xx}} \,
    \label{eq|slope}
\end{equation}
and
\begin{equation}
zp_{\rm y|x}=\langle y\rangle - m_{\rm y|x}\langle x\rangle \, ,
    \label{eq|zp}
\end{equation}
with the weighted averages $\langle y\rangle$ and $\langle x\rangle$. The other quantities are
\begin{equation}
S_{xx}=\sum_i \left(x_i-\langle x\rangle\right)^2 w_i \, ,
\quad
S_{yy}=\sum_i \left(y_i-\langle y\rangle\right)^2 w_i \, ,
    \label{eq|Sxx}
\end{equation}

\begin{equation}
S_{xy}=\sum_i \left(x_i-\langle x\rangle\right)\left(y_i-\langle y\rangle\right) w_i \, ,
    \label{eq|Sxy}
\end{equation}
\begin{equation}
E_{xx}=\sum_i \langle e_x^2\rangle_i w_i \, ,
\quad
E_{yy}=\sum_i \langle e_y^2\rangle_i w_i \, ,
    \label{eq|Exx}
\end{equation}
and
\begin{equation}
E_{xy}= \sum_i \langle e_x e_y\rangle_i w_i
     \approx k \sqrt{E_{xx}E_{yy}}\, .
    \label{eq|Exy}
\end{equation}
The terms $e_x$ and $e_y$ in \eq\ref{eq|Exx} represent the unknown measurement errors in the variables $x$ and $y$; only their variances $\langle e_x^2\rangle$ and $\langle e_y^2\rangle$ are known. The factor $k$ in \eq\ref{eq|Exy} accounts for correlation between the measurement errors $e_x$ and $e_y$. We will always set $k=0$ except when calculating the slopes and residuals in the \nb\ and \mbulge\ correlations, for which we set $k=0.9$ \citep{Meert13}, as the S\'{e}rsic index and galaxy luminosity are derived from the same fitting procedure.

In order to determine the final slope and correlation coefficient of the residual for each set of variables we proceed as follows. Suppose we have three variables, say, $x=\log \mbhe$, $y=\log \nbe$, and $z=\log \sigma$. We first calculate the correlation coefficient $r$ for each pair as
\begin{equation}
r_{\rm xy}=\frac{S_{xy}-E_{xy}}{\sqrt{S_{xx}-E_{xx}}\sqrt{S_{yy}-E_{yy}}} \,
    \label{eq|Rxy}
\end{equation}
and then compute the slope $m_{\rm xy|z}$ and correlation coefficient $r_{\rm xy|z}$ of the residual as
\begin{equation}
m_{\rm xy|z}=\frac{r_{\rm xy}-r_{\rm xz}r_{\rm yz}}{\left[1-r_{\rm yz}^2\right]}\sqrt{\frac{S_{xx}}{S_{yy}}} \, ,
    \label{eq|SlopeFinal}
\end{equation}
and
\begin{equation}
r_{\rm xy|z}=\frac{r_{\rm xy}-r_{\rm xz}r_{\rm yz}}{\sqrt{\left[1-r_{\rm xz}^2\right]\left[1-r_{\rm yz}^2\right]}} \, .
    \label{eq|CorrFinal}
\end{equation}
For each panel in \figus\ref{fig|Residuals} and \ref{fig|ResidualsLaesker} we ran 200 iterations following the steps outlined above and, in a bootstrap fashion, each time eliminating three objects at random from the original samples. From the full ensemble of realizations we then compute the mean slope of the correlation and its 1$\sigma$ uncertainty, which we report in the right, upper corner of each panel, while the upper left corner reports the mean value of the Pearson coefficient $r$. The analytic methodology described above is mainly intended for symmetric errors. To take into account the asymmetry in black hole mass uncertainties, for each correlation we ran 100 iterations considering only the positive error, and 100 iterations considering only the negative one. Considering instead the average or squared error in black hole mass yields consistent results within the uncertainties.

\section{Relation between coefficients in pairwise correlations and correlations between residuals}

The main text addresses the question of whether or not the \mbh-\nb\ correlation shown in \figu\ref{fig|Model} is fundamental.  We do so following \citet{ShethBernardi12}.   Namely, we start with \eq\ref{eq|MbhSigma} in the main text,
with $(\alpha,\beta)=(0.5,5)$, and around which there is 0.25~dex scatter that does not depend on \nb.

Averaging this expression over all \sis\ at fixed \mstar\ yields
\begin{equation}
 \langle\log\mbhe|\log\mstare\rangle \propto
 \alpha\,\log\mstare + \beta\,\langle\log\sise|\log\mstare\rangle\, .
 \label{eq|A1}
\end{equation}
If $\alpha_{\sise|*}$ is the slope of the $\langle\log\sise|\log\mstare\rangle$ relation, then we have that
\begin{equation}
 \langle\log\mbhe|\log\mstare\rangle \propto
 (\alpha + \beta\,\alpha_{\sise|*})\,\log\mstare\, ,
 \label{eq|A2}
\end{equation}
which suggests defining
\begin{equation}
 \alpha_{\rm tot} = \alpha + \beta\,\alpha_{\sise|*}\, .
 \label{eq|A3}
\end{equation}
Similarly, averaging over all \mstar\ at fixed \sis\ instead yields
\begin{equation}
 \beta_{\rm tot} = \beta + \alpha\, \beta_{*|\sise}\, ,
 \label{eq|A4}
\end{equation}
where $\beta_{*|\sise}$ is the slope of the $\langle\log\mstare|\log\sise\rangle$ relation.  This shows explicitly that $\alpha_{\rm tot}\ne\alpha$ and $\beta_{\rm tot}\ne\beta$, but that the relation between the two depends on the two projections of the \mstar-\sis\ correlation.  In our SDSS sample, $\sise\propto\mstare^{0.3}$ and $\mstare\propto\sise^{2}$, making $(\alpha_{\rm tot},\beta_{\rm tot})\approx (2,6)$ when $(\alpha,\beta)=(0.5,5)$. These values of $(\alpha_{\rm tot}$ and $\beta_{\rm tot})$ are in agreement with those reported in the left panels of \figus\ref{fig|Residuals} and \ref{fig|ResidualsLaesker} (long-dashed purple lines).

Of course, these relations should hold in the full sample:  selection effects may modify these relations and introduce curvature. This is indeed what we observe in the residuals at fixed S\'{e}rsic index (left panels of \figus\ref{fig|Residuals} and \ref{fig|ResidualsLaesker}). Our Monte Carlos, inclusive of the selection bias in the black hole's gravitational sphere of influence, predict significantly flatter, and in fact curved, residuals, roughly consistent with $(\alpha_{\rm tot},\beta_{\rm tot})\approx (1,4)$.

\begin{figure*}
    \center{\includegraphics[width=17truecm]{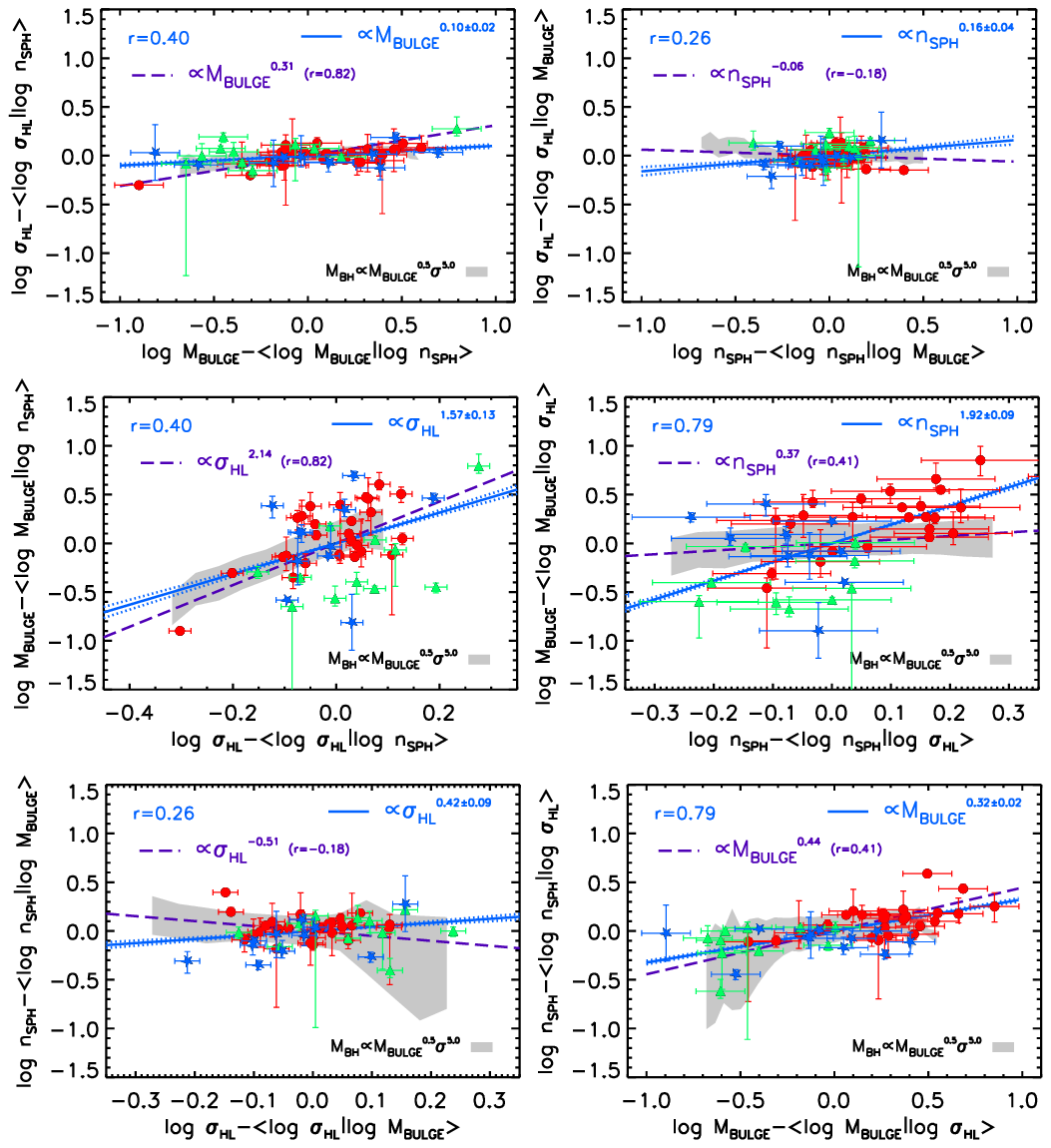}
    \caption{Residual correlations of velocity dispersion (top panels), bulge stellar mass (middle panels), and Sersic index (bottom panels) against the the other two variables. Gray bands are the results from the Monte Carlo simulations based on the \citet{Meert13} SDSS sample inclusive of bias on the black hole gravitational sphere of influence.
    \label{fig|AllOtherResidualsSavo}}}
\end{figure*}

\begin{figure*}
    \center{\includegraphics[width=17truecm]{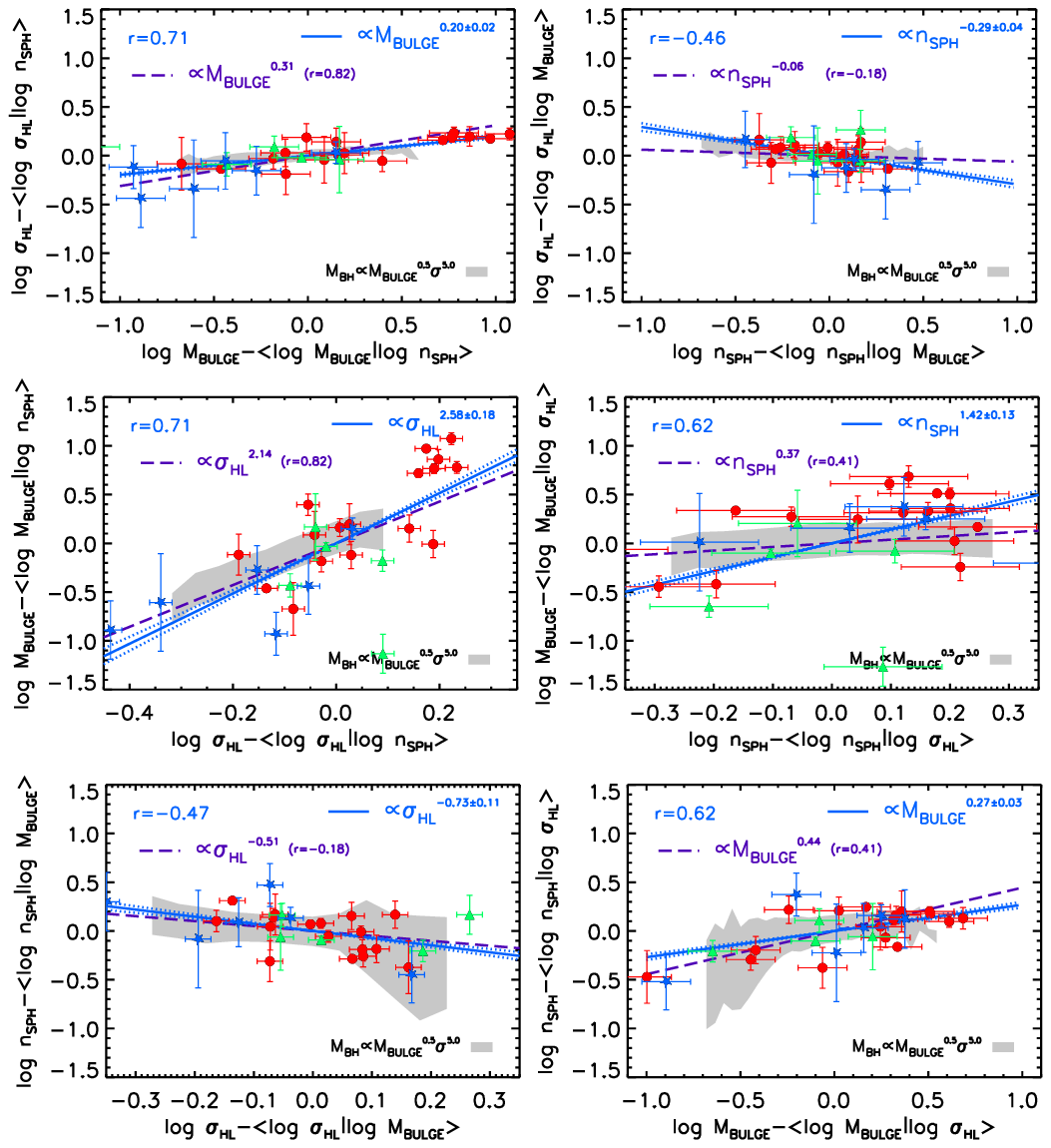}
    \caption{Same as \figu\ref{fig|Residuals} but for the \citet{Laesker14} sample. The residual correlations with S\'{e}rsic index at fixed velocity dispersion and stellar mass are again much weaker than those in the \citet{Savo15} sample.
    \label{fig|AllOtherResidualsLaesker}}}
\end{figure*}

Similarly, if the 0.25~dex scatter around \eq\ref{eq|MbhSigma} does not depend on $\nbe$, we expect correlations such as those in the top panels of \figus\ref{fig|Residuals} and \ref{fig|ResidualsLaesker} to satisfy
\begin{align}
 & \langle\log\mbhe|\log\nbe,\log\sise\rangle \nonumber\\
 &\propto \beta\log\sise\,  +
  \alpha\,\langle\log\mstare|\log\nbe,\log\sise\rangle\nonumber\\
 &\propto (\alpha\,\delta_{*|n\sise})\,\log\nbe\, +
          (\beta + \alpha\,\beta_{*|n\sise})\,\log\sise\, ,
          \label{eq|A5}
\end{align}
whereas those in the bottom panels should scale as
\begin{align}
 & \langle\log\mbhe|\log\nbe,\log\mstare\rangle \nonumber\\
 & \propto (\beta\,\delta_{\sise|n*})\,\log\nbe
 + (\beta\,\alpha_{\sise|n*} + \alpha)\,\log\mstare\, .
 \label{eq|A6}
\end{align}
These expressions show that, if the \mbh-\nb\ correlation is driven by the correlation between \sis\ and \mstar, and their correlations with \nb, then the coefficients of correlations between residuals depend both on the black-hole parameters $\alpha,\beta$, and on the \mstar-\sis-\nb\ correlations.  Specifically, in the top panels, the parameters which matter are those for $\mstare\propto \nbe^{\delta_{*|n\sise}}\sise^{\beta_{*|n\sise}}$, whereas it is $\sise\propto \nbe^{\delta_{\sise|n*}} \mstare^{\alpha_{\sise|n*}}$ which matters in the bottom panels.
Averaging \eq\ref{eq|A5} over \sis\ at fixed \nb\ yields
\begin{equation}
 \delta_{\rm tot} = \alpha\,(\delta_{*|n\sise} + \beta_{*|n\sise}\delta_{\sise|n})
                  + \beta\,\delta_{\sise|n},
 \label{eq|dtot1}
\end{equation}
and this equals the result of averaging \eq\ref{eq|A6} over \mstar\ at fixed \nb:
\begin{equation}
 \delta_{\rm tot} = \beta\,(\delta_{\sise|n*} + \alpha_{\sise|n*}\delta_{*|n}) + \alpha\,\delta_{*|n}.
 \label{eq|dtot2}
\end{equation}
These final expressions, show how the slope $\delta_{\rm tot}$ of the $\langle\log\mbhe|\log\nbe\rangle$ relation depends on the black-hole parameters $\alpha,\beta$, and on the scaling relations between \mstar, \nb\ and \sis.
The latter are reported in \figus\ref{fig|AllOtherResidualsSavo} and \ref{fig|AllOtherResidualsLaesker} for the \citet{Savo15} and \citet{Laesker14} samples, respectively. In each Figure the residual correlations of velocity dispersion (top panels), bulge stellar mass (middle panels), and S\'{e}rsic index (bottom panels) are plotted against the the other two variables. The gray band in each panel marks the results from the Monte Carlo simulations based on the \citet{Meert13} SDSS sample inclusive of bias on the black hole gravitational sphere of influence.

Inserting $\delta_{\sise|n}=0.16$ from Table~\ref{tab:SDSSslopes} in \eq\ref{eq|dtot1}, and the slopes of the SDSS residuals $\delta_{*|n\sise}=0.37$ and $\beta_{*|n\sise}=2.14$ from, respectively, the middle right and middle left panels of \figus\ref{fig|AllOtherResidualsSavo} and \ref{fig|AllOtherResidualsLaesker}, we would get $\delta_{\rm tot}=0.5(0.37+2.14\times 0.16)+5\times0.16 \approx 1.2$, implying a significant correlation between black hole mass and S\'{e}rsic index, even though \eq\ref{eq|MbhSigma} does not explicitly depend on S\'{e}rsic index. On the other hand, setting $\delta_{*|n}=0.36$ (Table~\ref{tab:SDSSslopes}), $\delta_{\sise|n*}=-0.06$ (upper right panels) and $\alpha_{\sise|n*}=0.31$ (upper left panels) in \eq\ref{eq|dtot2} yields $\delta_{\rm tot}=5(-0.06+0.31\times 0.36)+0.5\times0.36 \approx 0.5$.  This is weaker than the expected value of $1.2$; the discrepancy may be a consequence of the fact that $\delta_{\sise|n*}$ is so close to zero.

Except for this, all of the other self-consistency conditions are satisfied in the mocks before we apply the sphere of influence selection.  However, there is no guarantee that they will be satisfied in the selection-biased mocks or in the (selection-biased) data.

Nevertheless, the top panels of \figu\ref{fig|Residuals} suggest $\mbhe \propto \sise^{4.1}\nbe^{0.8}$ in the selection biased sample.  Using these values in \eq\ref{eq|dtot1}, along with the fact that $\delta_{\sise|n}\sim 0.3$ (left panel of \figu\ref{fig|SigmaSersic}) says that we expect $\delta_{\rm tot}\approx 0.8 + 4.1\,(0.3) \approx 2$.  This is close to the $\mbhe \propto \nbe^{1.8}$ we see in the left panel of \figu\ref{fig|Model}.  Using \eq\ref{eq|dtot2} instead means we should use the values in the bottom panels of \figu\ref{fig|Residuals} along with $\delta_{*|n}\approx 0.36$ (note that \figu\ref{fig|Sersic} shows the inverse relation, $\alpha_{n|*}$).  This yields $0.7 + 0.9\,(0.4)\approx 1.1$, which is somewhat lower than the slope of 1.8, perhaps again because the correlation with \nb\ is so weak. Since these scalings are satisfied in the full mocks, we conclude that these differences are due to the selection bias.

If we use the values in the top panel of \figu\ref{fig|ResidualsLaesker} instead, we find $0.4 + 3.9(-0.1) \approx 0.01$, where we have used the fact that $\langle\log\sise|\log \nbe\rangle\approx\sim -0.1$ for this sample (right panel of \figu\ref{fig|SigmaSersic}).  This is close to the $\mbhe \propto \nbe^{0.1}$ scaling of the direct relation shown in the right hand panel of \figu\ref{fig|Model}, despite the fact that this slope is very different from that in the left hand panel of \figu\ref{fig|Model}.  We conclude that these very different scalings are indicating that systematics in the determination of \nb\ prevent a definitive determination of some aspects of the \mbh-\nb-\sis\ relation.  However, the main uncertainties are related to the fact that correlations with \nb\ are not strong:  our finding that the \mbh-\sis\ correlation is stronger is very likely to be correct.

\footnotesize{
  \bibliographystyle{mn2e_Daly}
  \bibliography{../RefMajor_Rossella}
}

\label{lastpage}
\end{document}